\documentclass[11pt]{article}
\usepackage{latexsym}
\usepackage{times}
\usepackage{amsfonts} 
\newtheorem{Thm}{Theorem}
\newtheorem{Lem}{Lemma}
\newtheorem{Def}{Defintion}
\newtheorem{Prop}{Propostion}
\begin{document}
\title{Maximum weight spectrum codes with reduced length}
\author{G\'{e}rard D.\ Cohen and Ludo Tolhuizen \thanks{G.D.\ Cohen (deceased) was with T\'{e}l\'{e}com Paris, CRNS LTCI; 
L.\ Tolhuizen  (ludo.tolhuizen@philips.com) is with Philips Research, Eindhoven}}
\date{}
\maketitle
\begin{abstract}
A $q$-ary linear  code of dimension $k$  is called a maximum weight spectrum (MWS) code if it has the
maximum possible number ({\em viz.} $(q^k-1)/(q-1)$) of  different non-zero weights.
We construct MWS codes from quasi-minimal codes, thus obtaining of much shorter length than hitherto known.
 By an averaging argument,
we show the existence of MWS codes of even shorter length.
\end{abstract}
\section{Introduction}
In a recent preprint \cite{Shi} Shi {\em et al.} introduced and studied $L(k,q)$, the maximum number of non-zero weights that a linear code 
of dimension $k$ over $\mathbb{F}_q$ can have.
 As two non-zero linearly dependent words have equal weight, it is clear that \cite[Proposition 2]{Shi}
\begin{equation}\label{eq:basic}
L(k,q) \leq \frac{q^k-1}{q-1}.
\end{equation}
Shi {\em et al.} provide constructions showing that equality holds in (\ref{eq:basic}) for $q=2$ and all $k$, and for $k=2$ and all prime powers $q$.
They conjectured that equality holds in (\ref{eq:basic}) holds for all $k$ and all prime powers $q$.
Shortly afterwards, Alderson and Neri showed this conjecture to be true \cite{Ald}, and  called $q$-ary $k$-dimensional codes with exactly
$(q^k-1)/(q-1)$ non-zero words Maximum Weight Spectrum codes, or MWS codes for short. Alderson and Neri provide two code constructions, and
lower bounds on the length of MWS codes.
Another proof of the conjecture of Shi {\em et al} was provided in \cite{Men}.

The present note continues the study of MWS codes. More particularly, we construct  MWS codes from quasi-minimal codes \cite{Coh},
and thus obtain MWS  codes of much shorter length than hitherto known. Finally, by an averaging argument, we show the existence of MWS codes of even shorter length.

The second author dedicates this manuscript to the memory of G\'{e}rard Cohen, who sadly passed away during 
its preparation.
\section{Notation and definitions}
In this section, we establish notation and provide some defintions.

For prime powers $q$, and positive integers $n$ and $k$, a $k$-dimensional code over $\mathbb{F}_q$ of length $n$ and minimum Hamming distance
 $d$ is denoted as an $[n,k,d]_q$ code.
Similarly, a  $k$-dimensional code over $\mathbb{F}_q$ of length $n$ is denoted as an $[n,k]_q$ code.
For $x\in\mathbb{F}_q^n$, the support of $x$, denoted by supp$(x)$, is the set of indices of the non-zero entries of $x$. That is,
 supp($x$) =  supp$(x_1,\ldots, x_n)  = \{ i\mid x_i\neq 0\}$.
\begin{Def}\label{QM}
A  code $C$ is called {\em quasi-minimal} (QM) if any pair of linearly indepedent codewords have different supports.
That is, if $c,c'\in C$ are linearly independent, then supp$(c) \neq $supp$(c')$,
\end{Def}
Quasi-minimal codes were introduced and studied in \cite{Coh}. Obviously, the $[n,n]_2$ code is quasi-minimal.
\begin{Def}\label{MWS}
A code $C$ is called {\em Maximum Weight Spectrum} (MWS) if any pair of linearly independent codewords have different weights.
That is, if $c,c'\in C$ are linearly independent, then wt$(c) )\neq \mbox{wt}(c').$
\end{Def}
\begin{Prop}
Any MWS code is a QM code.
\end{Prop}
{\bf Proof} Obvious. 
\begin{Def}
For any property $P$, we define $n_{P,k,q}$ as the shortest length of a $q$-ary code of dimension $k$ that has property $P$.
That is,
\begin{equation} 
n_{P,k,q} = \min\{ n\in \mathbb{N} \mid \mbox{ there exists an } [n,k]_q \mbox{ code  with property } P \}
\end{equation}
\end{Def}
The constructions of \cite[Sec.\ 3] {Ald} and \cite[Sec.\ 4,5]{Ald} imply that 
\begin{equation}
 n_{MWS,k,q} \leq 2^{\frac{q^k-1}{q-1}-1} \mbox { and } n_{MWS,k,q} \leq {\cal O}(q^{k(k+1)-4}) . 
\end{equation}
Moreover, as shown in \cite[Lemma 5.1]{Ald},
\begin{equation}\label{eq:lowbound}
 n_{MWS,k,q} \geq \frac{q}{2}\cdot \frac{q^k-1}{q-1}.
\end{equation}
By combining  \cite[Thm.1]{Shi} and
(\ref{eq:lowbound}), it follows that
\begin{equation}
n_{MWS,k,2} = 2^k-1.
\end{equation}
As shown in \cite{Ald},  we have that
\begin{equation}
n_{MWS,2,q} = \frac{1}{2}q(q+1). 
\end{equation}
\section{Explicit construction of MWS codes from QM codes}\label{sec:geometric}
For each $n\geq 1$ and each prime power $q$, we define the embedding function $f:\mathbb{F}_q^n \rightarrow \mathbb{F}_q^{2^n-1}$ as 
\begin{equation}
f(c_0,c_1,\ldots, c_{n-1}) = (c_0,c_1,c_1,c_2,c_2,c_2,c_2,\ldots, \overbrace{c_i,c_i,\ldots, c_i}^{2^i \mbox{\small{times}}}, \ldots, c_{n-1},..,c_{n-1}).
\end{equation}
\begin{Prop}\label{Prop:QMtoMWS}
If $C$ is an $[n,k]_q$ QM code, then $f(C)$ is an $[2^n-1,k]_q$ MWS code.
\end{Prop}
{\bf Proof}
It is clear that for each $c\in\mathbb{F}_q^n$, we have that
\begin{equation}\label{eq:weight} \mbox{wt}\left(f(c)\right) = \sum_{i\in\mbox{\small{supp}}(c)} 2^i.
\end{equation}
Equation~\ref{eq:weight} implies that there is a one-to-one relation between the support of a code word $c$ and the weight of $f(c).
\;\;\;\Box$ \\ \\
We remark that $f(C)$ is a a special case of what is called a "generalized $r$-repetition of $C$" in \cite{Ald}.
\begin{Prop}\label{prop:lowbound}
For all $k$ and prime powers $q$, we have that $n_{MWS,k,q}\leq 2^{n_{QM,k,q}}$.
\end{Prop}
{\bf Proof} Direct consequence of Proposition~\ref{Prop:QMtoMWS}. $\;\;\;\Box$ \\ \\
We can take for $C$  an $[n,k_q]$ code for which all non-zero words have equal weight, for example, 
an $[n_k=(q^{k}-1)/(q-1),k,q^{k-1}]_q$ code, the  dual of the $[n_k,n_k-k,3]_q$ Hamming code. This choice is equivalent to 
what is done in \cite[Sec.\ 3]{Ald}, and results in a  $[2^{(q^k-1)/(q-1)-1},k]_q$  MWS code. \\
If we take for $C$ an $[k,k]_2$ code, we obtain a $[2^k-1,k]_2$ MWS code. This is in fact the explicit construction from \cite[Thm. 1]{Shi}. 

We now apply results from \cite{Coh} concerning QM codes.
It is shown in \cite{Coh} that an $[n,k,d]_q$ code is quasi-minimal whenever $\frac{d}{n}> \frac{q-2}{q-1}$. Combining this with 
the Gilbert-Varshamov bound, it follows
\cite{Coh} that there exists a quasi-minimal  $[n,k]_q$ code whenever $\frac{k}{n} \leq 1-h_q((q-2)/(q-1))$, where 
$h_q$ is the $q$-ary 
entropy function,
\begin{equation} h_q(x) = -x\log_q(x) -(1-x)\log_q(1-x) + x\log_q(q-1). 
\end{equation}
In other words, we have that
\begin{equation}
n_{QM,k,q} \leq k\cdot \lambda_q \mbox{ where } \lambda_q = (1-h_q(\frac{q-2}{q-1}))^{-1}. 
\end{equation}
 We conclude that for all inegers $k$ and prime powers $q$,
\begin{equation} n_{MWS,q,k} \leq 2^{k\lambda_q}. 
\end{equation}
By direct computation,  it can be found that 
$\lambda_q =  2q^3\log q(1+{\cal O}(q^{-1}))$, where $\log$ denotes the natural logarithm.
In the next section, we provide a construction of MWS codes of smaller length.
\\ \\ \\
{\bf Remark} We can obtain sharper upper bounds on $n_{MWS,q,k}$ by using sharper bounds for $n_{QM,k,q}$. \\
First, it can be shown\cite{AsBa} that an $[n,k]_q$ code with minimum distance $d$ and maximum weight $D$ is QM whenever $\frac{d}{D} > \frac{q-2}{q-1}$.
Combining this with a generalization of GV we obtain a QM code of improved rate. Unfortunately, we  do not have a closed form expression for the rate,
it involves is a root of an algebraic equation involving the $q$-ary entropy function. \\
Alternatively, by a small adaptation of the proof of \cite[Thm. 2]{Coh}, it can be shown that
\[ n_{QM,k,q} \leq  k \mu_q \mbox{ with } \mu_q=2 / \log_q(\frac{q^2}{q^2-2q+2}). \]
This seems to yield the shortest length for QM codes but is "totally" non-constructive.
Note that  $\mu_q= q\log q(1+{\cal O}(q^{-1}))$, so  for large $q$, $\mu_q$ is approximately $2q^2$ as small as $\lambda_q$.

\section{A non-constructive bound upper bound to $n_{MWS,k,q}$ }\label{sec:randomcode}
\begin{Thm}\label{Thm:4kexistence}
For each prime power $q$, there exists an integer $D_q$ such that $n_{MWS,k,q}\leq D_q\cdot q^{4k}$ for all $k\geq 1$.
\end{Thm}
In order to prove this theorem, we use  several lemmas.
Throughout this section,  we denote the number of words of weight $w$ of a code $C$ by $A_w(C)$.
\begin{Lem}\label{lem:suff} A $q$-ary code $C$  is an MWS code if and only if
\begin{equation}\label{eq:wtdis}  \sum_{w=1}^n A_w(C) (A_w(C)-(q-1)) \;\; < 2 (q-1)^2. 
\end{equation}
\end{Lem}
{\bf Proof} If $C$ is an MWS code, then $A_w(C)\in\{0,q-1\}$ for each $w\geq 1$, so (\ref{eq:wtdis}) is satisfied.
If $C$ is not an MWS code, then $A_w(C)\geq 2(q-1)$ for at least one $w. \;\;\;\Box$
\begin{Lem}
Let $n,k$ be integers and let $q$ be a power of a prime.
If
\begin{equation}\label{eq:bound}
 q^{2k-2n} \sum_{w=0}^n {n\choose w}^2 (q-1)^{2w} < 2 (q-1)^2, 
\end{equation}
then there is a $q$-ary $[n,k]$ MWS code.
\end{Lem}
{\bf Proof}  We denote by ${\cal C}$ the set of all $q$-ary $[n,k]$ codes.
Each pair of linearly independent vectors in $\mathbb{F}_q^n$  clearly is in equally many, say $\lambda$, codes in ${\cal C}$.
By counting in two ways triples $x,y,C$ with $C\in{\cal C}$ and $x,y$ linearly independent vectors in $C$, we infer that
\begin{equation}\label{eq:lambda}
(q^n-1)(q^n-q)\lambda = |{\cal C}| (q^k-1)(q^k-q).
\end{equation}
By counting in two ways the number of pairs $(x,y,C)$ with $C\in{\cal C}$ and $x,y$ linearly independent codewords of
 equal weight in $C$,
 we infer that
\begin{equation}
\sum_{C\in{\cal C}} \sum_{w\neq 0} A_w(C)(A_w(C)-(q-1)) =
\sum_{w\neq 0} {n\choose w}(q-1)^w \left( {n\choose w}(q-1)^w-(q-1) \right) \lambda. 
\end{equation}
As a consequence,
\begin{equation} \frac{1}{|{\cal C}|} \sum_{C\in{\cal C}} \sum_{w\neq 0} A_w(C)(A_w(C)-(q-1)) < 
\frac{\lambda}{|{\cal C}|} \sum_w {n\choose w}^2(q-1)^{2w}. 
\end{equation}
Equation~ (\ref{eq:lambda}) implies that $\frac{\lambda}{|{\cal C}|} \leq q^{2k-2n}$, and hence,
 by using (\ref{eq:bound}),  we infer that 
\begin{equation} 
\frac{1}{|{\cal C}|} \sum_{C\in{\cal C}} \sum_{w\neq 0} A_w(C)(A_w(C)-(q-1)) < 2(q-1)^2. 
\end{equation}
As a consequence, there is a $C\in{\cal C}$ such that $\sum_{w\neq 0} A_w(C)(A_w(C)-(q-1)) < 2(q-1)^2$.
According to Lemma~\ref{lem:suff}, $C$ is MWS. $\;\;\;\Box$ 

\begin{Lem}
For each integer $q$, there is constant $C_q$ such that for each positive integer $n$ 
\[ \sum_{w=0}^{n} {n\choose w}^2 (q-1)^{2w} < C_q \frac{q^{2n}}{\sqrt{n}} . \]
\end{Lem}
{\bf Proof} (sketch) For positive integers $n$ and $q$, let
\begin{equation} M(n,q) = \max\{ {n\choose w} (q-1)^w \mid 0\leq w\leq n\}.
\end{equation}
We obviously have that 
\begin{equation}
 \sum_{w=0}^n {n\choose w}^2(q-1)^{2w} \leq \sum_{w=0}^n M(n,q) \cdot {n\choose w}(q-1)^w = M(n,q)q^n. 
\end{equation}
As
\[ \frac{ {n\choose w}(q-1)^{w}}{{n\choose w-1}(q-1)^{w-1}} = \frac{(q-1)(n-w+1)}{w}, \]
it follows that ${n\choose w}(q-1)^{w}\geq {n\choose w-1}(q-1)^{w-1}$ if and only if  $w\leq\frac{q-1}{q}(n+1)$.
This implies that ${n\choose w}(q-1)^w$ is maximal for $w=w_{max}=\lfloor\frac{q-1}{q}(n+1)\rfloor$, so
$w_{max}\approx \frac{q-1}{q}n$.
Application of Stirling's formula yields that 
\[  \log {n\choose \frac{q-1}{q}n} (q-1)^n \approx n\log q + \frac{1}{2}\log\left( \frac{q^2}{q-1}\frac{1}{n}\right) -\log(\sqrt{2\pi}), \]
so that $M(n,q)\approx q^n \cdot \sqrt{\frac{q^2}{q-1}\cdot \frac{1}{n}}. \;\;\;\Box$
\\ \\
By  combining the above lemmas, we obtain Theorem~\ref{Thm:4kexistence} (with $D_q= C_q/(2(q-1)^2) \approx
\frac{q}{2(q-1)^{5/2}})$.

\section{Conclusions and open problems}
By using a construction of MWS codes from QM codes, combined with the Gilbert-Varshamov bound, we obtained
MWS codes of length $n=2^{k\lambda_q}$,
  where $\lambda_q$ grows as $2q^3\log q$. 
We provided a non-constructive existence proof  of  $[n,k]_q$ MWS codes of length $n=D_q q^{4k}$, where $D_q$ is 
independent of $k$. \\
Combining the existence bound with (\ref{eq:lowbound}), we obtain that
\begin{equation}\label{eq:ineq} \frac{q}{2}\cdot \frac{q^k-1}{q-1} \leq n_{MWS,k,q} \leq D_q q^{4k} . 
\end{equation}
It is an interesting problem to reduce the gap in the above inequality.
Another interesting problem is to determine, for a fixed $q$, the exisence or non-existence of the limit
\[ \lim_{k\rightarrow \infty} \frac{1}{k} \log_q (n_{MWS,k,q}).  \]
Note that (\ref{eq:ineq}) implies that for each prime power $q$
\[  \liminf_{k\rightarrow\infty} \frac{1}{k} \log _q(n_{MWS,k,q}) \geq 1
\mbox{ and } \limsup_{k\rightarrow\infty} \frac{1}{k} \log_q(n_{MWS,k,q})\leq 4 . \]
We finally remark that in \cite{Men}, Meneghetti studies the related problem of the existence
 of $[n,k]_q$ codes with an arbitrary number of weights.
He proves the existence for the binary case, and conjectures it for the $q$-ary case.

\end{document}